\documentclass[a4paper,fleqn]{cas-sc}

\usepackage[numbers]{natbib}

\newcommand{\eg}{\textit{e.g.}}

\begin{document}
\let\WriteBookmarks\relax
\def\floatpagepagefraction{1}
\def\textpagefraction{.001}

\shorttitle{The Impact of ChatGPT and LLMs on Medical Imaging Stakeholders}

\shortauthors{Jiancheng Yang et~al.}

\title [mode = title]{The Impact of ChatGPT and LLMs on Medical Imaging Stakeholders: Perspectives and Use Cases}                      

\nonumnote{This work was supported in part by the Swiss National Science Foundation.}

\author[1]{Jiancheng Yang}[orcid=0000-0003-4455-7145]

\cormark[1]

\cortext[cor1]{Corresponding author}


\ead{jiancheng.yang@epfl.ch}

\ead[url]{jiancheng-yang.com}

\credit{Conceptualization, Methodology, Investigation, Writing - Original Draft, Writing - Review \& Editing}

\affiliation[1]{organization={Computer Vision Laboratory, Swiss Federal Institute of Technology Lausanne (EPFL)},
    city={Lausanne},
    citysep={}, 
    postcode={1015}, 
    country={Switzerland}}

\author[2,3]{Hongwei Bran Li}[orcid=0000-0002-5328-6407]
\ead{hongwei.li@tum.de}
\ead[url]{hongweilibran.github.io}

\credit{Conceptualization, Writing - Review \& Editing}

\affiliation[2]{organization={Technical University of Munich},
    city={Munich},
    citysep={}, 
    postcode={85748}, 
    country={Germany}}

\affiliation[3]{organization={University of Zurich},
    city={Zurich},
    citysep={}, 
    postcode={8001}, 
    country={Switzerland}}

\author[4]{Donglai Wei}[orcid=0000-0002-2329-5484]
\ead{donglai.wei@bc.edu}
\ead[url]{donglaiw.github.io}

\credit{Conceptualization, Writing - Review \& Editing}

\affiliation[4]{organization={Boston College},
    city={Chestnut Hill},
    citysep={}, 
    postcode={02467}, 
    state={MA},
    country={USA}}

\begin{abstract}
This study investigates the transformative potential of Large Language Models (LLMs), such as OpenAI ChatGPT, in medical imaging. With the aid of public data, these models, which possess remarkable language understanding and generation capabilities, are augmenting the interpretive skills of radiologists, enhancing patient-physician communication, and streamlining clinical workflows. The paper introduces an analytic framework for presenting the complex interactions between LLMs and the broader ecosystem of medical imaging stakeholders, including businesses, insurance entities, governments, research institutions, and hospitals (nicknamed \emph{BIGR-H}). Through detailed analyses, illustrative use cases, and discussions on the broader implications and future directions, this perspective seeks to raise discussion in strategic planning and decision-making in the era of AI-enabled healthcare.
\end{abstract}

\begin{keywords}
ChatGPT \sep LLM \sep foundation models \sep medical imaging
\end{keywords}

\maketitle

\section{Introduction}

Artificial Intelligence (AI) has been at the forefront of revolutionary changes in numerous sectors, and healthcare is no exception~\cite{gulshan2016development,jin2020deep,ouyang2020video,moor2023foundation}. At the heart of these changes lie foundation models~\cite{bommasani2021opportunities}, an emergent class of AI technologies known for their large-capacity and generalizability nature. Their learning process involves encoding patterns and relationships on a vast array of data, often spanning diverse domains of human knowledge. This allows them to acquire a rich, although implicitly, understanding of the concept relation and world knowledge. Among foundation models, Large Language Models (LLMs), such as OpenAI's GPT-3~\cite{brown2020language}, ChatGPT, and the more recent GPT-4~\cite{bubeck2023sparks}, are designed to predict the next word in a sequence. This task requires a broad understanding of syntax, semantics, and even world facts. LLMs are trained on diverse Internet text, thereby inheriting a broad understanding of human language. The emergent abilities~\cite{weiemergent} of these models extend beyond simple information retrieval; they can generate creative content, answer complex questions,  translate languages, and could reasonably be viewed as an early version of an artificial general intelligence (AGI) system~\cite{bubeck2023sparks}.
This adaptability is integral to their broad applicability across various disciplines, including healthcare.

In the healthcare sector, the burgeoning availability of open data~\cite{antonelli2022medical,bilic2023liver,yang2023medmnist,johnson2023mimic} is enhancing the accessibility and effectiveness of foundation models in healthcare~\cite{singhal2023towards}. Integrating foundation models creates transformative opportunities in healthcare, notably in healthcare data interpretation, disease diagnosis and prognosis, patient treatment and management~\cite{lee2023benefits,moor2023foundation,haupt2023ai}. LLMs, particularly ChatGPT and GPT-4, recently have shown promise in medical imaging, aiding in text-related tasks like interpreting radiology reports~\cite{bhayana2023performance} and structured reporting~\cite{adams2023leveraging}.

Medical imaging is a cornerstone of modern healthcare~\cite{varoquaux2022machine}. The integration of LLMs in this field can augment the interpretive skills of radiologists, facilitate patient-physician communication, and streamline workflows in clinical settings, usually happening in hospitals. However, the impact of these advancements is not limited to the clinical aspects of healthcare; it permeates throughout the broader ecosystem of medical imaging stakeholders. This paper proposes an analytic framework named \emph{BIGR-H}---a crafted acronym hinting at the notion of a ``Bigger Hospital''---to investigate the impact of LLMs on medical imaging stakeholders. The BIGR-H stands for businesses (\textbf{B}), insurance entities (\textbf{I}), governments (\textbf{G}), research institutions (\textbf{R}), and hospitals (\textbf{H}), which provides a structured approach to dissect the complex interactions between LLMs and various stakeholders in the medical imaging landscape. 

In the subsequent sections, we elaborate on the impacts of LLMs on each BIGR-H stakeholder, present several illustrative use cases, and delve into the broader implications and future directions of LLM integration in the medical imaging domain. By providing a comprehensive understanding of these elements, this perspective aims to contribute to the strategic planning and decision-making processes in the era of AI-enabled healthcare.

\section{Impacts on Medical Imaging Stakeholders (BIGR-H)}

The healthcare ecosystem is an intricate and dynamic network designed to address multiple layers of health needs, from prevention and primary care to specialized and palliative care. This complex, interconnected network of various entities is orchestrated around the key mission of delivering effective healthcare services to all individuals. This intricate network of relationships motivates us to introduce the \emph{BIGR-H} framework for discussion, encompassing businesses (\textbf{B}), insurance entities (\textbf{I}), governments (\textbf{G}), research institutions (\textbf{R}), and hospitals (\textbf{H}). Each of these stakeholders brings unique values and objectives to the ecosystem. Businesses, including biopharmaceutical and medical device companies, fuel innovation and bring new therapies and technologies to the market. Insurance entities play a critical role in risk pooling and financing healthcare services. Governments, on the other hand, are primarily responsible for regulation, health policy formulation, and ensuring that the population's health needs are met. Research institutions generate scientific knowledge and insights that propel healthcare innovation, while hospitals and other healthcare providers deliver essential healthcare services directly to patients. The interactions among these stakeholders are often multi-faceted and highly interdependent.

In this complex web of interactions, LLMs such as ChatGPT are emerging as influential players. With their unparalleled ability to interpret health data, improve communication, and facilitate decision-making processes, these models are transforming the roles and operations of each stakeholder in the healthcare ecosystem~\cite{lee2023benefits,moor2023foundation}. Thus, the integration of LLMs into the healthcare ecosystem is not just enhancing existing capabilities but also redefining the interactions and balance among the stakeholders, heralding a new era of high-performance medicine~\cite{topol2019high}---a paradigm defined by technology-enabled, data-driven healthcare services.

In the subsequent section, we will examine the wide-ranging effects of ChatGPT and LLMs on various medical imaging stakeholders.

\paragraph{Businesses (\textbf{B}).}

Within the medical imaging ecosystem, businesses play a critical role in bringing technological advancements to market. Medical device manufacturers often pioneer cutting-edge diagnostic equipment, while contract research organizations (CROs) expedite the development of new devices and treatments through clinical trials. These entities handle vast data volumes, making them ideal candidates for LLM integration.

Medical device companies are vital components in the evolution of healthcare by developing sophisticated equipment capable of capturing intricate details of the human body. Some have already incorporated AI into their offerings, crafting medical imaging tools that provide more accurate diagnoses and predictions~\cite{benjamens2020state}. LLMs can serve as valuable tools for analyzing copious amounts of user feedback and technical documentation, which could provide significant insights, leading to the evolution of existing devices. The applicability of LLMs can also optimize user manuals and device operation guidelines into talking agents, enhancing their clarity and comprehensibility for healthcare professionals. Besides, existing AI diagnostic tools augmented by LLMs can facilitate interactive patient consultation, thereby improving patient understanding and engagement.

In parallel, CROs could harness the power of LLMs to re-engineer their data processing and analysis workflows in clinical trials. Traditionally, clinical trial data management is labor-intensive and often prone to human error. By employing LLMs in data handling and analysis, these organizations can bolster the speed and accuracy of their research processes. The resultant streamlining could lead to shorter turnaround times and more reliable outcomes, thereby accelerating the translation of research into viable treatments.

\paragraph{Insurance (\textbf{I}).}

Insurance entities, including health insurance companies and public health insurance providers, serve as vital intermediaries, assuming the financial risk of healthcare expenses on behalf of patients and making healthcare services financially accessible to the public. They stand to gain significantly from the integration of LLMs into their operations. These models can enhance customer service, facilitate more accurate risk assessment, and improve fraud detection. With the ability to process large volumes of data, LLMs can identify patterns and anomalies that could signify potential fraudulent activities, providing an essential tool for insurance companies to prevent such occurrences. In addition, LLMs can aid in answering policyholder queries and providing personalized advice, improving the overall customer experience and ensuring policyholders receive the most accurate and helpful information.

\paragraph{Government (\textbf{G}).}
Government entities, encompassing regulatory agencies (\eg{}, NMPA and FDA), along with public health authorities (\eg{}, health commissions and centers for disease control and prevention), play a fundamental role in ensuring the safety and efficacy of medical products and overseeing public health initiatives.

The integration of LLMs could significantly streamline the operations of these entities. For regulatory agencies, LLMs can augment the regulatory review process, assisting in the meticulous scrutiny of medical product submissions. On the other hand, public health authorities could also leverage the analytical prowess of these models to scrutinize large volumes of health data. By identifying trends and patterns, LLMs could significantly enhance disease surveillance capabilities, providing valuable insights for proactive disease control and prevention measures. Moreover, these insights could inform the formulation of more targeted and effective health policies, optimizing resource allocation and contributing to overall public health outcomes. 

\paragraph{Research (\textbf{R}).}
Academic institutions and industry research stand on the brink of a transformative leap with the integration of LLMs into their research endeavors. These models are adept at interpreting and analyzing extensive biomedical datasets, facilitating more precise conclusions and novel discoveries. In addition, their capacity to discern research gaps and generate new hypotheses can catalyze medical research and innovation.

In bridging the gap between research and education, particularly in medical imaging, LLMs hold substantial promise. They could function as personalized instructors, elucidating complex imaging concepts and results, thereby fostering a more intuitive understanding for students. By creating a learning environment that simulates real-world imaging scenarios and provides interactive interpretations, LLMs can enrich the learning experience.

\paragraph{Hospital (\textbf{H}).}

The delivery of clinical services takes place in various settings, but hospitals, encompassing a broad spectrum of medical examination institutions such as radiology service providers and physical examination centers, are at the core. These settings are integral parts of the healthcare service delivery chain where LLMs can have profound implications, significantly affecting the dynamics of medical imaging processes.

Within the scope of medical imaging in hospitals, stakeholders such as radiologists, referring physicians, hospital administrators, and patients can interact uniquely with LLMs. Radiologists and referring physicians can utilize LLMs as advanced decision-support tools. These models' prowess in interpreting vast quantities of radiology reports can potentially augment diagnostic precision and streamline imaging processes. 
Hospital administrators, often tasked with coordinating various hospital units, could incorporate LLMs to enhance administrative tasks such as scheduling and inter-departmental communication. 
On the other hand, patients, as the central figures in the healthcare landscape, stand to benefit from LLMs serving as advanced communication platforms. These models can distill complex medical information into more understandable terms, encouraging patients to participate actively in their healthcare journey.

\section{Use Cases}

In this section, we explore various use cases of ChatGPT and LLMs in medical imaging, illustrating their potential to enhance clinical workflows, consultations, and medical education and research. The examples are generated via prompt engineering OpenAI ChatGPT.

\paragraph{Optimizing Clinical Workflow.} 

\begin{figure}
\centering
\includegraphics[width=\linewidth]{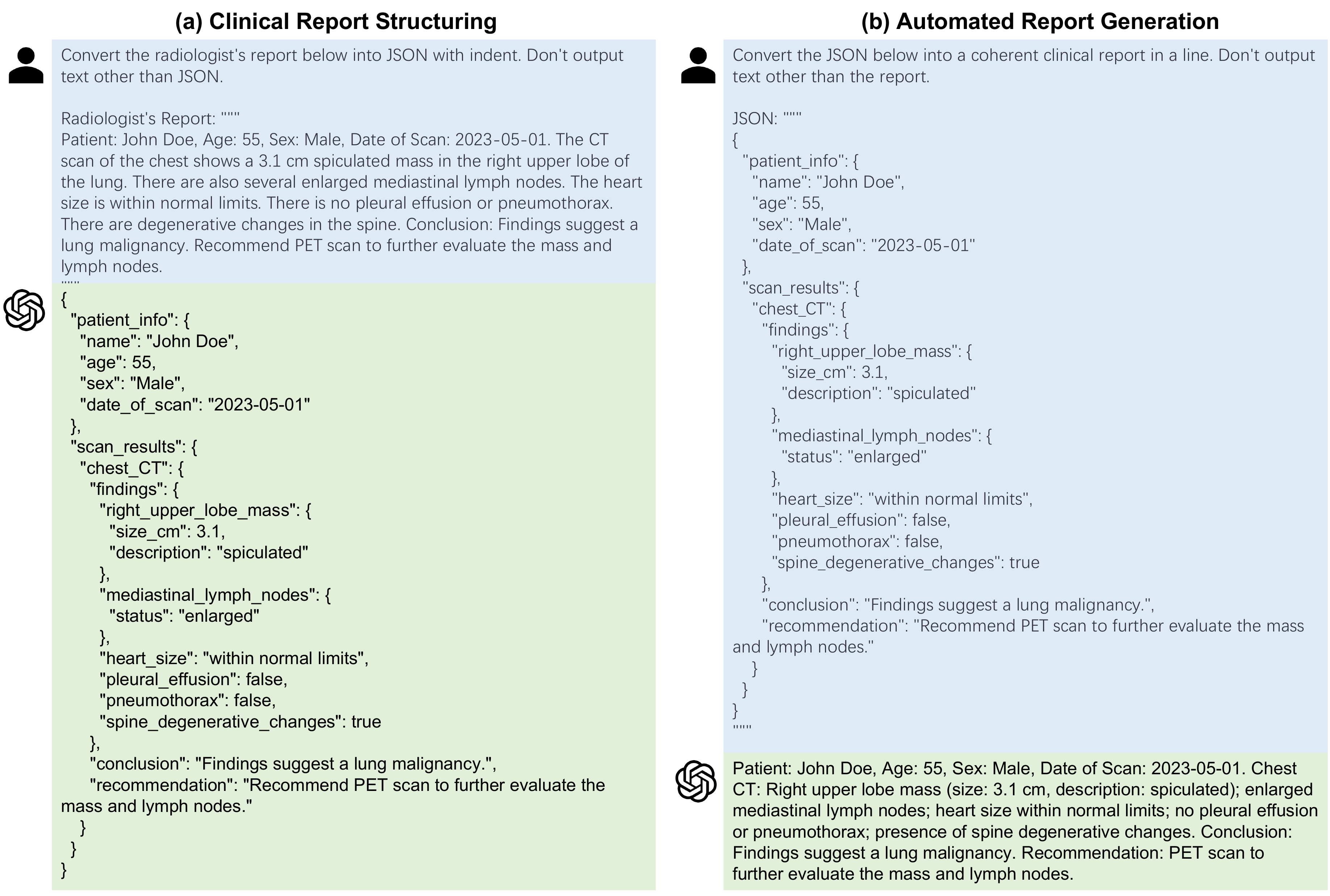}
\caption{\textbf{Optimizing Clinical Workflow.} (a) Clinical Report Structuring. (b) Automated Report Generation.}
\label{fig:json_text}
\end{figure}

LLMs have the potential to significantly optimize clinical workflows, particularly in medical imaging contexts. Through their ability to interpret and analyze vast quantities of data quickly and accurately, they can streamline several time-consuming tasks in clinical workflows~\cite{adams2023leveraging}. 
One critical application is in the structuring and generation of clinical reports. For example, interpreting scans and writing reports in radiology can be time-consuming. As illustrated in Fig.~\ref{fig:json_text}, LLMs could help automate part of this process by transforming unstructured text into structured data (clinical report structuring) and vice versa (automated report generation). 

In clinical report structuring, LLMs can convert a narrative radiology report into a structured format such as JSON, which is easier to process and analyze. This not only improves data management but also enhances interoperability across healthcare systems. On the other hand, in automated report generation, LLMs can utilize structured clinical data (\eg{}, JSON), which could be generated by another AI systems, to create a narrative report that can be easily understood by physicians and patients alike. This application can potentially reduce the administrative burden on healthcare professionals, allowing them to focus more on direct patient care.

\paragraph{Enhancing Patient Communication.}

\begin{figure}
\centering
\includegraphics[width=\linewidth]{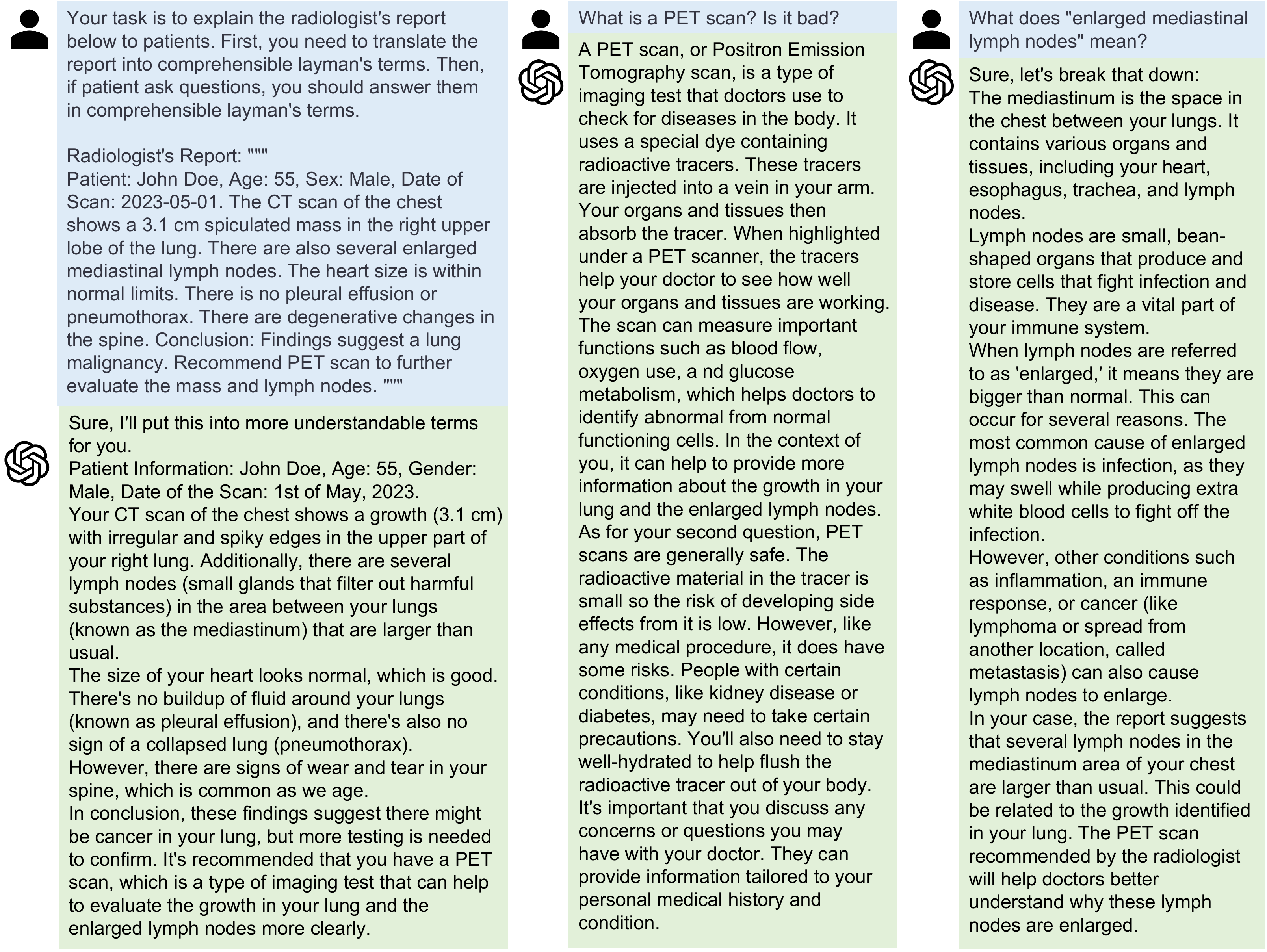}
\caption{\textbf{Enhancing Patient Communication,} where an interactive Q\&A is illustrated given a clinical report.}
\label{fig:interactive_diagnosis}
\end{figure}

LLMs can serve as powerful tools to enhance patient communication. Often, radiology reports and medical jargon can be perplexing for patients, creating a barrier to understanding their own health status. LLMs can bridge this gap by translating complex medical language into comprehensible layman's terms, thereby improving patient understanding and engagement~\cite{ayers2023comparing}. For instance, given a radiology report, LLMs can generate a simplified explanation and answer patient queries about their diagnosis in a natural language conversation (Fig.~\ref{fig:interactive_diagnosis}). The interactive diagnosis provided by LLMs, combined with their capacity to handle numerous and diverse patient inquiries, marks an advancement in personalized patient communication.

\paragraph{Supporting Medical Education and Research}

\begin{figure}
\centering
\includegraphics[width=\linewidth]{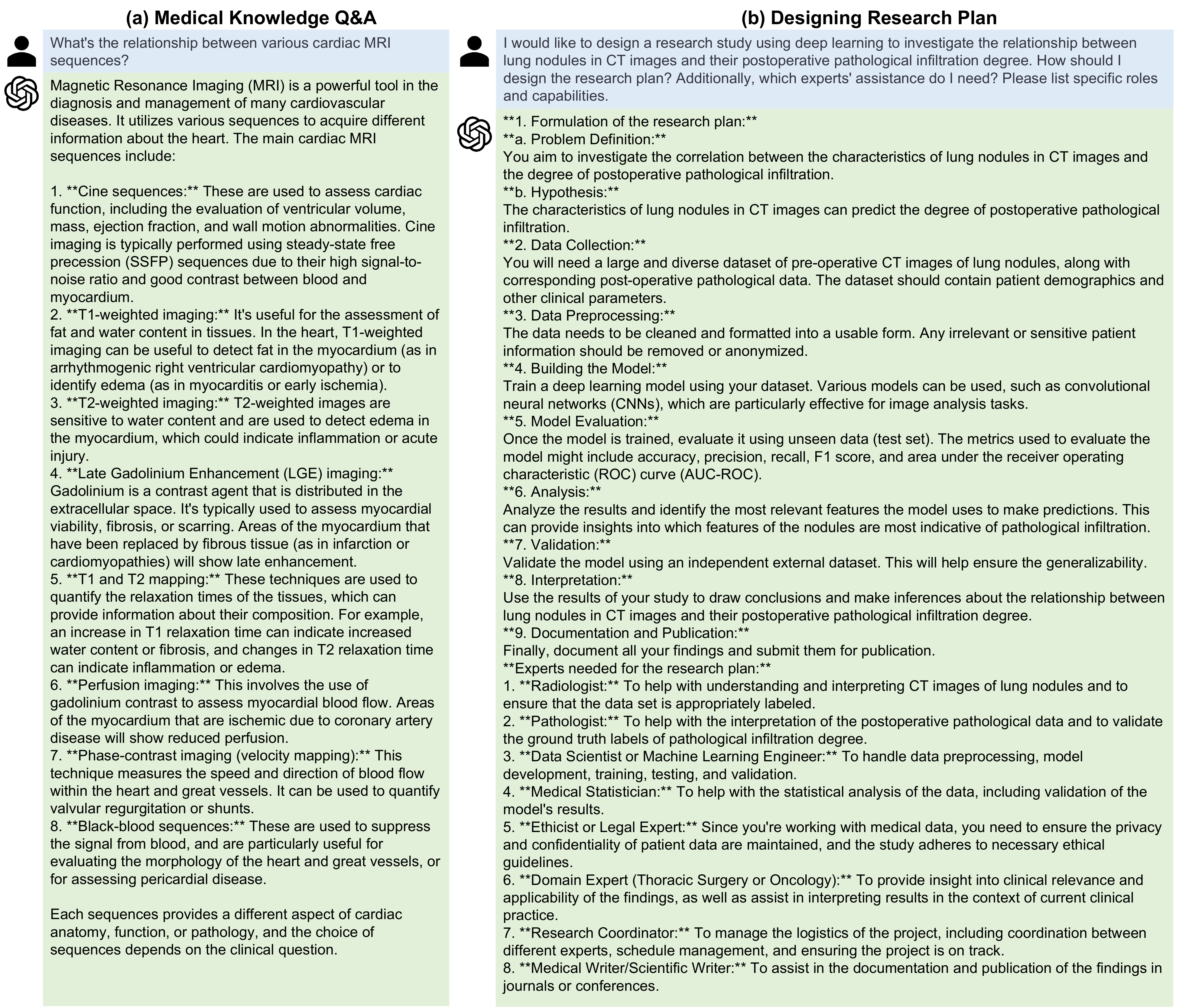}
\caption{\textbf{Supporting Medical Education and Research.} (a) Medical Knowledge Q\&A. (b) Designing Research Plan.}
\label{fig:edu_research}
\end{figure}

LLMs are becoming increasingly impactful in the realm of medical education and research, notably in the specialized field of medical imaging. In the educational setting, LLMs can serve as virtual tutors~\cite{lee2023benefits}, answering complex questions about different imaging techniques or specificities of certain pathological findings. For example, a student can ask about the relationship between various cardiac MRI sequences (Fig.~\ref{fig:edu_research} (a)), and the LLM could provide a detailed, yet understandable explanation. This personalized learning assistance can supplement traditional teaching methods and support students in navigating complex medical subjects.

In terms of research, LLMs can assist in the design of research plans by providing insights on experimental design, data analysis methods, or even suggesting relevant literature. As illustrated in Fig.~\ref{fig:edu_research} (b), given a research problem, an LLM can help conceptualize the research structure, formulate hypotheses, and propose suitable methods for data collection and analysis. Notably, the answer aligns well with previous studies for the same research questions~\cite{zhao20183d,yang2020hierarchical}.

\section{Implications and Future Directions}

The incorporation of LLMs like ChatGPT in medical imaging has begun to profoundly reshape this field, presenting immense opportunities and transformative potential for all involved parties. The advantages offered by these models are twofold. Firstly, they bring an extraordinary capability to decode complex language structures, leading to simplified understanding and generation of diverse and intricate medical jargon. This novel approach heralds a new era of communication and data interpretation, essentially making the discourse in medical imaging more accessible and meaningful to a broader audience---from seasoned clinicians and radiologists to healthcare staff and patients. Secondly, LLMs empower entities with logical programming abilities traditionally reserved for computer scientists and software engineers. By making these capabilities more accessible, LLMs democratize advanced technology, thereby providing an equal opportunity for all healthcare stakeholders to contribute to and benefit from AI's progress.

These advantages manifest as tangible improvements in healthcare delivery. LLMs can improve diagnostic precision by providing more nuanced and comprehensive interpretations of medical imaging data. Efficiency in clinical workflows is enhanced as LLMs automate the reading and interpretation of diagnostic imaging reports, reducing manual workload and allowing healthcare professionals to devote more time to patient care. Furthermore, by generating patient-friendly explanations of imaging results, LLMs can enhance patient engagement, making healthcare a more collaborative and holistic experience.
Another major advantage of LLMs is their potential to revolutionize medical education. They can provide personalized, intuitive, and interactive learning experiences, adapting to individual learning styles and pacing. This dynamic environment fosters a deeper understanding of complex medical imaging concepts and enhances skill development.

Despite the considerable benefits, the advent of LLMs in medical imaging raises several significant challenges~\cite{lee2023benefits,shen2023chatgpt}. Ethical issues abound, primarily concerning the degree of trust and responsibility placed in AI technology. It's crucial to contemplate the potential ramifications when AI errs or fails to deliver the expected results. Alongside this, the safety and privacy of patient data are of paramount importance. LLMs deal with vast amounts of sensitive patient data, making robust security measures and strict adherence to privacy regulations non-negotiable.

Bias and fairness are also critical issues~\cite{zou2018ai,ricci2022addressing}. LLMs learn from the data they are trained on; if this data is inherently biased, these biases could be replicated in AI predictions, leading to unfair outcomes and exacerbating existing healthcare disparities. Regulatory and legal challenges also emerge, as the integration of LLMs requires clear, comprehensive guidelines from regulatory bodies to protect patient safety and ensure data integrity.

Looking ahead, future directions for LLMs in medical imaging could include exploring multi-modality~\cite{bubeck2023sparks,moor2023foundation}, wherein LLMs are trained on a combination of text, medical imaging, and other omics data~\cite{karczewski2018integrative,yang2021multi,ding2022improving} for more nuanced analysis and interpretation. Improving liability and reducing malpractice is another crucial aspect~\cite{gabriel2020artificial}. As LLMs become increasingly integrated into patient care, systems must be put in place to deal with potential errors and their repercussions.

In conclusion, the deployment of LLMs like ChatGPT offers an exciting new frontier in medical imaging, yielding promising use cases that can revolutionize clinical workflows, patient communication, and medical education and research. Despite the considerable challenges, with continued research, collaboration, and careful consideration of ethical and regulatory guidelines, the future of LLMs in medical imaging holds remarkable promise.

\printcredits

\end{document}